\newcommand{\beq}{\begin{equation}}
\newcommand{\eeq}{\end{equation}}
\newcommand{\beqa}{\begin{eqnarray}}
\newcommand{\eeqa}{\end{eqnarray}}

\renewcommand{\lambda}{\ell}
\renewcommand{\vec}[1]{{\bf #1}}
\documentstyle[aps,prl,twocolumn,epsf]{revtex}
\begin{document}
\def\dfrac#1#2{{\displaystyle{#1\over#2}}}

\twocolumn[\hsize\textwidth\columnwidth\hsize\csname @twocolumnfalse\endcsname

\title{On the Lifshitz tail in the density of states of a
  superconductor with magnetic impurities}

\author{
  A.V.~Balatsky$^{\ast}$ and  S. A. Trugman
}
\address{Theoretical Division, Los Alamos National Laboratory, Los Alamos, 
New Mexico 87545}
  
\date{\today}
\maketitle

\begin{abstract}
 We argue that {\it any} 
superconductor with magnetic impurities is gapless due to a
Lifshitz tail in the density of states extending to zero energy. At low
   energy the density of states   
$\nu(E \rightarrow 0)$ remains finite. We show that
   fluctuations in the impurity distribution produce regions of
   suppressed superconductivity, which are responsible for the low
   energy density of states.

\


\end{abstract}

\

\

]


The role of impurities in superconductors is a rich subject, going
 back to the pioneering papers by Abrikosov and Gor'kov \cite{AG} and
 by Anderson \cite{And}.  However the majority of work has been
 concentrated so far on the ``mean field" treatment of the impurity problem
 in superconductors. Here we will address the role of the fluctuations
 of the distribution of magnetic impurities in an s-wave
 superconductor.

It has been experimentally known for some time that the density of
states (DOS) in a superconductor with magnetic impurities is far greater
 at low energies than one would expect from Abrikosov-Gor'kov theory
\cite{tunnel}.  Using the suppression of the critical temperature to infer
the  pairbreaking parameter, one typically arrives at a substantially
lower DOS at $E\ll \Delta_0$ than is observed.
($\Delta_0$ is the superconducting gap in
the spectrum.)  We suggest here that the {\em observed
deviations from the Abrikosov-Gor'kov theory at small $E$ are caused by
fluctuations in the impurity distribution and the Lifshitz tail in the DOS of
an impure superconductor}.

We observe that for any, no matter how small, concentration $n$ of
magnetic impurities in a superconductor there are fluctuations in the
distribution of impurities across the sample.  There are finite 
 regions of high impurity concentration, where the 
superconducting state is suppressed due to scattering.  These large
regions of essentially normal metal produce low lying, $E \ll \Delta_0$ 
 single particle
states in the averaged density of states of the superconductor. It
is clear that any singularity in the DOS, if one occurs, should be at $E=0$
due to the particle-hole symmetry of the superconducting state, which we assume
here and which is preserved even with magnetic impurities.  We
find that at low energy $E\ll \Delta_0$ the DOS is:
\begin{equation}
\nu(E) \propto (1/\Delta_{L_0}) exp(-const ~ L_0^{~d})  ,
 \end{equation}
where   $\nu(E)$ is scaled with 
the normal state DOS, $d$ is the dimensionality of space,
 $\Delta_{L_0}$ is the mean
 level spacing in the fluctuation region of the size of
 length $L_0 = (\xi_0 l)^{1/2}$, where
 $\xi_0 = \pi v_f/\Delta_0$ is the $T=0$ superconducting
 coherence 
length, and $l$ is the mean free path. The constant in the exponent will be given below. The tail in the 
DOS of a superconductor is
similar to the tail in the DOS of a semiconductor, the so-called
Lifshitz tail \cite{Lif}.

For any particular model of impurity scattering (e.g. Born versus unitary
 scattering), we assume that  there
 exists  a critical concentration $n_c$ at which a
 thermodynamic superconducting sample will become normal due to the
 pairbreaking effect of impurities.  The specific value of $n_c$ obviously
 depends on the model.  For  the case of the Born scattering limit of
 magnetic impurities, within  the Abrikosov-Gor'kov theory,  
$n_c = O(1) \Delta_0 N_0 /(J^2 N_0^2 S(S+1))$, 
where $N_0$ is the normal
 metal DOS, $J$ is the magnetic exchange between conduction electrons and
the impurity, and $S$ is the magnitude of the impurity spin \cite{AG}. This
 specific value is not important for our subsequent considerations. We
 will use $n_c$ as a model--dependent input to our final answer. All
 concentrations are given in terms of the dimensionless concentration per
 unit cell of linear size $a$.

\begin{figure}
\epsfxsize=2.2in
\centerline{\epsfbox{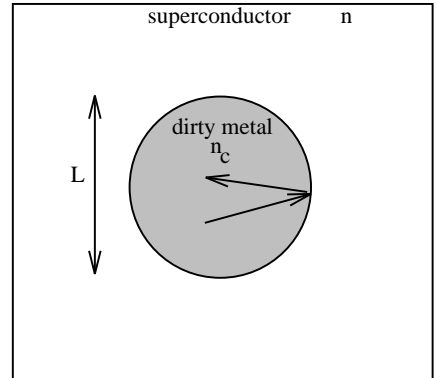}}
 \caption[]{\small Fluctuation region of size $L$,
 with concentration of impurities $n_c$ inside the superconductor,
 is shown schematically. The equilibrium
 concentration is $n < n_c$. 
The fluctuation region has a metallic spectrum.
  Andreev reflection modifies the spectrum of quasiparticles \cite{mix}. In any 
local probe of the DOS, e.g. an STM, one would find that the $I-V$ 
characteristics have a gap in the outer region, but are gapless 
if measured at any point inside the fluctuation region. 
The average DOS of a
superconductor as $E \rightarrow 0$ hence will be 
the sample average of the DOS of the
fluctuation regions.}
\end{figure}

Here  we will 
consider the case of arbitrary  impurity exchange strength. 
It is known that magnetic
 impurities induce intra-gap   states \cite{YSR}. 
The energy of these states for large $S$ is approximately
$\omega_0 = \Delta_0 (1 - J^2 N_0^2 S(S+1)) / (1+J^2 N_0^2 S(S+1))$.
  These impurity   states have a wavefunction
 $\Psi(r)\sim exp(-r/\xi_{\omega_0})$ of 
size $\xi_{\omega_0} = \xi_0 (1-(\omega_0/\Delta_0)^2)^{-1/2} \geq \xi_0$, 
where $\xi_0$ is the zero temperature superconducting 
coherence length. For subsequent consideration we assume that 
$n_c (\xi_0/a)^d \gg 1$, generally true for 
realistic systems, so that 
intra-gap 
states are strongly overlapping in the region where the impurity 
concentration is $n_c$.
 Impurity states form an impurity band,
 centered around $\omega_0$ \cite{YSR}. Fluctuations in the 
distribution of impurities lead to tails in this impurity
 band, which extend to zero energy.

Consider a fluctuation in the impurity distribution
 such that inside a region $V(L) = L^d$ \cite{comment5}, the 
local concentration of impurities is $n_c$ 
(averaged over distances much greater than
 $\xi_{\omega_0}$ but smaller than $L$), as
 shown in Fig. 1. We assume that
  $L \gg \l \geq \xi_{\omega_0}$, where
 $l$ is the mean free path at the critical impurity 
concentration $n_c$.

The  low energy single particle spectrum in the
 fluctuation region will be  normal, since the local
 concentration is $n_c$. The proximity coupling 
to the superconducting reservoir at the boundary cannot
open up a gap at large distances $\sim L \gg \xi_0$ 
due to pairbreaking
scattering.
 We ignore the region of size $\xi_0$ from the boundary where the
 gap is decaying. The single particle spectrum inside $V(L)$ 
 will be equivalent to the spectrum of a normal metallic 
region with magnetic impurities in tunneling contact with a bulk
 superconductor.

To verify  that the  spectrum of
the fluctuation region  is indeed gapless we  
have numerically calculated
 the spectrum of a random superconductor in the mean field
 approximation.
Specifically, we considered the 1D BCS superconductor with the
 Hamiltonian
\begin{eqnarray}
 H = -t \sum_{<i,j>, \sigma}
 c^{\dag}_{i, \sigma}c_{j, \sigma} +
 \sum_{i}\Delta^*_{i}c_{i, \uparrow}c_{i,\downarrow} + h.c. + \nonumber\\
J 
\sum_{i \in V(L),\alpha,\beta} {\vec {S}}_i \cdot 
c^{\dag}_{i,\alpha}{\vec{\sigma}}_{\alpha,\beta}c_{i,\beta}
\label{randomham}
\end{eqnarray} 
where  $i$ labels the sites of 1D chain, $V(L)$  
are the impurity sites,
$t$ is the nearest-neighbor electron hopping,
 $\Delta_i$ is the pairing amplitude on the site
$i$, $J$ is the exchange coupling between the conduction 
electron and impurity spin,
and  ${\vec{S}}_i$ is a {\em random} classical
Heisenberg impurity spin on the site $i$.
The last term in Eq. (\ref{randomham}) describes the impurity scattering 
effects of the fluctuation region,
which we assume 
to be in the middle of  the superconducting region.

We consider a superconducting system
of 40 sites with impurity spins present at a high
concentration $x$ on 10 of these sites.
This approximation was chosen 
to mimic the high impurity 
density fluctuation region, which is responsible for the 
low energy DOS.  For classical spins, the coupling $J$ 
and  impurity spin magnitude enter 
into the answer in the combination $JS$, 
hence the specific values of each of them separately
does not matter.  We have
calculated the spectrum of quasiparticles in the mean 
field approximation, ignoring the
self-consistency condition for the gap \cite{selfconsistency}.
The DOS for this model is shown in Fig. (2).

\begin{figure}
\epsfxsize=2.5in
\centerline{\epsfbox{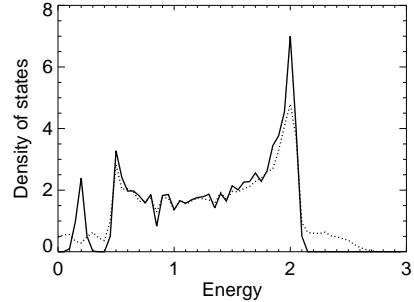}}
 \caption[]{\small  The density of states is plotted for a 1D BCS
superconductor with 40 sites.  
There is an impurity region
of 10 sites, in which the superconducting gap
$\Delta$ is taken to be zero.  Classical Heisenberg
impurity spins with  random orientation
occupy the impurity region with concentration $x=0.5$.
The solid line is for coupling $JS=0.1$, and the dotted
line is $JS=1.0$.  Other parameters are $t=1$ and $\Delta = 0.5$.
The DOS is averaged over 1500 realizations.
The fine-scale roughness in the middle of the band
is due to finite size effects.
}
\end{figure}

Since $\Delta = 0$ in the impurity region, there
are intra-gap states even for $JS=0$.  (There is
only one such state for the parameters of Fig. 2.)
We find that 
the intra-gap state evolves into 
an impurity band, 
and gradually fills the entire gap as the concentration 
or the coupling constant $JS$ increases. This evolution
of the impurity band is similar to the evolution of the
band in doped semiconductors.  
The calculation confirms all
the basic features one might expect: the
appearance of impurity states inside the gap region, 
the growth of the impurity band, and finally the filling of
states at low energies with nonzero  $\nu(0)$.

\begin{figure}
\epsfxsize=2.5in
\centerline{\epsfbox{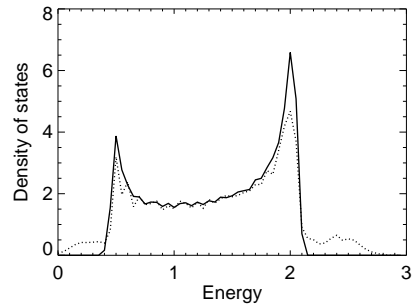}}
  \caption[]{\small   Same as figure (2), but with the superconducting gap
$\Delta = 0.5$ throughout the sample, including
the impurity region.
}
\end{figure}

A similar calculation is shown in Fig. (3), but 
with the mean-field superconducting gap $\Delta = 0.5$ everywhere,
including the impurity region.  In this case,
there are no intra-gap states for $JS=0$.
For small $JS$, intra-gap states first
appear at $\omega_0$, which is just below
the energy $\Delta$ of the uniform gap.  
As $JS$ increases, the gap gradually fills in.
A larger value of $JS$ is required to 
completely close the gap
than in Fig. (2), because
the density must spread down from $\Delta$ rather
than from the intra-gap levels that already exist 
in Fig. (2) at $JS=0$.

A similar problem for a metallic grain in the presense of time
reversal violating fields (e.g. impurity spins) in contact with 
a superconductor was considered by Altland and
 Zirnbauer \cite{AZ}. At energies small compared to the Thouless
 energy $E_T = D/L^2$, one can ignore the spatially inhomogeneous
 solutions of the nonlinear-$\sigma$-model. In this limit the spectrum
 of the grain is given by random matrix theory.
 The single particle DOS in Ref.\cite{AZ} is $\nu_L(E) = 1/\Delta_L \left(1 +
  {sin(2\pi E/\Delta_L)\over 2\pi E/\Delta_L}\right)$, which goes to constant as $E \rightarrow 0$. Here $\Delta_L$ (not to be confused with the gap $\Delta_0$) 
is the mean level spacing of the grain of linear size $L$, and
$\nu_L(E)$ is averaged over all realizations of the random
 spectrum for grains of size $L$.  We are 
interested in $E \ll \omega_0$, where the constraint $E\ll E_T$ is
 not important since
  $E_T/\omega_0 \sim (\Delta_0/\omega_0) (\xi_0 l/L^2)$ and 
  this ratio is small except in the limit $\omega_0 \rightarrow 0$, 
where special care should be taken. We will not  address this 
limit here.  We believe that 
 the result $\nu(E) \sim const$ will still hold.

If we make the assumption that the spectrum of the fluctuation
 region in Fig. 1 is equivalent to the spectrum of a normal metal
 grain, it is easy to estimate the average DOS 
$\nu(E)$ from the distribution $P_L(n_c;n)$ 
in the size of the fluctuation
 regions:
\begin{eqnarray}
\nu(E)  \sim \int dV(L) P_L(n_c;n) \nu_L(E) .
\label{average}
\end{eqnarray}

We now consider the probability distribution for a normal region of
volume $V(L)$ with linear size $L \gg \xi_0$ to occur.  This question
is equivalent to finding the probability $P_L(n_c;n)$ of a
fluctuation region of diameter $L$, taken to be spherical in
d-dimensions, with a concentration of impurities in this region equal
to or greater than $n_c$, while the average concentration is $n$.
This probability can be easily evaluated, following, for example, the
arguments of Refs. \cite{Lif,ES}. We find:
\begin{eqnarray}
log (P_L(n_c;n)) = \delta \sigma = -V(L)\phi(n_c;n)\nonumber\\
\phi(n_c;n) \simeq n_c  log(n_c/n) -n_c+n , 
\label{prob}
\end{eqnarray}
 where $\delta \sigma$ is the change in entropy due to a 
fluctuation with homogeneous concentration $n_c$ in the region $V(L)$, and
$\phi(n_c;n)$ is the entropy density for the discussed fluctuation,
which is model dependent. 
Equation (\ref{prob}) applies for small $n$ and $n_c$ \cite{Gd}.
 Strictly speaking, Eq. (\ref{prob}) gives the probability of
a fluctuation with a concentration equal to $n_c$.
 In principle one should
integrate this probability over the range $n \geq n_c$
to obtain the total probability that
the normal region $V(L)$ will occur.  Taking
into account this effect will only change the coefficient in
$\phi(n_c;n)$ and the prefactor in Eq. (1).

The ratio of the mean level spacing to the superconducting
gap is given by
$\Delta_L/\Delta_0 = \kappa(n_c, J, N_0) L^{-d}$,
 where $\kappa(n_c, J, N_0)$ is a model--dependent dimensionless
 function of $n_c, J, N_0$ \cite{comment3}.  With the aid of Eq. (\ref{prob}) 
and
 using $\Delta_L$ we find:
\begin{eqnarray}
\nu(E) =   \int_{V(L_0)} dV(L) P_L(n_c;n) \nu_L(E) \nonumber\\
\sim \Delta^{-1}_{L_0}  exp(-L_0^{~d} \phi(n_c;n)) 
\label{final}
\end{eqnarray}
This is our main result.
  In 
writing
Eq. (\ref{final}) the lower limit of the volume integration was taken
at $L = L_0 = (\xi_0 l)^{1/2}$, when $E_T \sim \Delta_0$,
 because at smaller distances the gap acquires a nonzero mean 
value
 due to strong coupling to the bulk 
superconductor.

The fact that $\nu(E)$ is nonzero at arbitrarily small energy 
implies  that a superconductor with magnetic impurities is always
gapless. This does not, however, mean that the system is not
superconducting.  A DC current can flow through the system 
(around the impurity regions) with no
dissipation, i.e., there is a condensate.  The dissipation is nonzero
for essentially any AC current due to dissipation in the normal metal regions.

A few comments are in order here.  i) It should be noted that there is
a qualitative difference between the DOS in the tails for a
superconductor as compared to a  semiconductor.  In the case of a high impurity
concentration region of size $L$ in a semiconductor, the energy has a
quadratic dispersion $E - E_0 = \pi^2/2mL^2$, where $E_0$ is the
lowest energy of the crystal composed of only the impurity atoms. This
results in a $\nu(E) \propto exp(-const/(E-E_0)^{d/2})$ for the Lifshitz tail in
a semiconductor. The difference comes from the fact that in a semiconductor
tails are formed near the bottom of the band, whereas in our case the
destruction of the superconducting gap leads to disordered normal
regions.  ii) The suppression of superconductivity occurs at quite a low
concentration $n_c \sim 1\%$. This allows for substantial fluctuations
of the impurity distribution inside $V(L)$. However it is clear that
the most important configuration responsible for the low lying states
is the one with a nearly homogeneous distribution with local
concentration close to $n_c$.  Any fluctuations with local $n(r)
\leq n_c$ are ineffective for $\nu(E \rightarrow 0)$. 
We expect any improvement of the above consideration will lead to
corrections to $\phi(n_c;n)$. iii)  We have ignored the
possible interactions between impurity spins.  This does
not have to be the case in real systems, where in order to suppress
superconductivity one has to have many impurities in regions of the
size of the coherence length: $n_c (\xi_0/a)^d \gg 1$.  Interactions
between spins in this situation may be important, as was pointed out
by Larkin et. al. \cite{LMK}.  iv) Similar considerations should
apply to any system with a spontaneously induced gap due to
interactions, i.e. CDW, SDW  systems and to unconventional, e.g. d-wave, superconductors.

The present work is related to that of Larkin and Ovchinnikov
\cite{LO}. They considered the DOS fluctuations for a  disordered
superconductor due to fluctuations of the gap $\Delta_0 ( {\bf r} )$,
and also found that the DOS is finite at small energies
due to this process. We have considered here a different mechanism for
generating a nonzero DOS.

We benefited from discussions with Markku Salkola at an early 
stage  of this work. We are grateful to B.L. Altshuler, A.F. Andreev, A.R. 
Bishop,  A.I. Larkin, D.J. Scalapino and M. Zirnbauer for useful
discussions. We are grateful to G. Aeppli for bringing Refs.(\cite{tunnel}) to our attention. Part of this work was done at ITP, Santa
Barbara.  This work has been supported by the Department of Energy and by 
NSF Grant PHY94-07194 at ITP.


\begin{references}

\item[$^*$] Also at the Landau Institute for Theoretical Physics,
 Moscow, Russia.


\bibitem{AG} A.A. Abrikosov and L.P. Gor'kov, Soviet Phys. JETP {\bf 12},1243 
(1961).


\bibitem{And} P.W. Anderson, Phys. Rev. Lett. {\bf 3}, 325 (1959);
T. Tsuneto, Prog. Theor. Phys. {\bf 28}, 857 (1962);
D. Markowitz and L.P. Kadanoff, Phys. Rev. {\bf 131}, 563 (1963).

\bibitem{tunnel} See for example: M.A. Wolf and F. Reif, Phys. Rev. {\bf 137}, A557 (1965); A.S. Edelstein, Phys. Rev. Lett. {\bf 19}, 1184 (1967); S.D. Bader, et.al., Solid State Communications, {\bf 16}, 1263 (1975). 





\bibitem{Lif} I.M. Lifshitz, Soviet Physics Uspekhi, 
{\bf 7}, 549, (1965); (Uspekhi Fiz. Nauk, {\bf 83}, 617, (1964)).

\bibitem{mix} This particle-hole admixture will enter in terms 
of ``coherence factors'', 
determining the effective charge of the excitation and related to
the coefficients in Eq. (1). This effect is only one of
the sources of uncertainty in this coefficient and will not be
addressed in detail here.  

\bibitem{YSR} L. Yu, Physica Sinica {\bf 21}, 75, (1965); H.
 Shiba, Progr. Theor. Phys. {\bf 40}, 435, (1968); 
A.I. Rusinov, Sov. Phys. JETP Letters {\bf 9}, 85, (1969).

\bibitem{comment5} We ignore constants in the relation 
 between V(L) and L as the answer will be given up to 
an unknown constant. 

\bibitem{selfconsistency} For the spectrum of a region of size 
$L \gg \xi_0 ,$ the self-consistent solution will presumably modify the 
gap only in a small region of size $\xi_0$ near the surface. The 
statistical properties of the spectrum of the 
fluctuation region will not be affected by 
self-consistent modifications of the gap function. 

 

\bibitem{AZ} A. Altland and M. Zirnbauer, cond-mat/9602137.



\bibitem{ES} B.I. Shklovskii and A.L. Efros, ``Electronic properties of doped 
semiconductors", p. 277, ch 12.3. Springer Verlag, (1984).

\bibitem{Gd} Recall that for superconducting
  metals  it takes only a few percent of magnetic
 impurities to suppress superconductivity completely,
 e.g., superconducting La with 1\% Gd becomes a normal metal.


\bibitem{comment3} The specific dependence of $\Delta_L$
in e.g. the unitary
versus the Born scattering limit 
on $n_c$ and $JS$ is determined by the particular
 model. A few features, however, remain
universal regardless of the strength of the impurity 
scattering and concentration.  The total number of  impurity
 generated states inside the region $V(L)$ is $N(L) = n_cV(L)$. 
The bandwidth
of the impurity band at small concentrations is 
$W \propto n^{1/2}_c ,$ and one finds qualitatively
$\Delta_L = W/N(L) \propto L^{-d}$ with all the 
model--dependent coefficients being assembled in $\kappa(n_c, J, N_0)$.

 


\bibitem{LMK} A.I. Larkin, V.I. Melnikov and D.E. Khmelnitskii, ZhETP, {\bf 60}, 
846, (1971),(Sov. Phys. JETP, {\bf 33}, 458, (1971)).


\bibitem{LO} A.I. Larkin and Y.N. Ovchinnikov, ZhETP, {\bf 61}, 2147, (1971), 
(Sov. Phys. JETP, {\bf 34}, 1144, (1972)).
  
\end{references}
\end{document}